\documentclass[pra,twocolumn,amssymb] {revtex4}
\usepackage{graphicx}

\begin{document}

\title{Double-Layer Bose-Einstein Condensates with Large Number of Vortices}
\author{
Hui Zhai, Qi Zhou, Rong L$\ddot{u}$ and Lee Chang,
\\ \it{Center for Advanced Study, Tsinghua University, Beijing,
100084, China}
 }
\date{\today}
\begin{abstract}
In this paper we systematically study the double layer vortex
lattice model, which is proposed to illustrate the interplay
between the physics of a fast rotating Bose-Einstein condensate
and the macroscopic quantum tunnelling. The phase diagram of the
system is obtained. We find that under certain conditions the
system will exhibit one novel phase transition, which is
consequence of competition between inter-layer coherent hopping
and inter-layer density-density interaction. In one phase the
vortices in one layer coincide with those in the other layer. And
in another phase two sets of vortex lattices are staggered, and as
a result the quantum tunnelling between two layers is suppressed.
To obtain the phase diagram we use two kinds of mean field
theories which are quantum Hall mean field and Thomas-Fermi mean
field. Two different criteria for the transition taking place are
obtained respectively, which reveals some fundamental differences
between these two mean field states. The sliding mode excitation
is also discussed.

\end{abstract}
\maketitle

\section{Introduction}

In the recent years remarkable progress has been made in the field
of ultracold quantum gas, among which the achievement of fast
rotating boson gases and the realization of Mott insulator to
superfluid transition are two important ones. The physics of these
two phenomena have attracted a lot of theoretical and experimental
interests.

It has been long predicted that there exists a quantum phase
transition from the superfluid phase to the Mott insulator phase
in the boson Hubbard model as the hopping parameter decreases, and
recently it has been observed in ultracold Bose atoms in the
optical lattice.\cite{Bloch} The Mott insulator phase is
characterized by the loss of phase coherence among the sites and
the whole system becoming fragmented. The essential physics of
this transition can be demonstrated by a simpler model, namely the
bosons confined in a double well potential. In the strong
tunnelling limit, the condensates located in the two wells develop
a relative phase and the ground state is described by a coherent
state. In the opposite limit, the relative number fluctuation is
suppressed and each condensate has determined atom
number.\cite{Legget} The ground state is well described by a Fock
state. Hence there exists a crossover from the coherent regime to
the Fock regime as the parameter $t/U$ decreasing in the two-well
system.

On the other hand, the boson atoms confined in the a
quasi-two-dimensional harmonic trap can now be rotated fast enough
that the rotation frequency is very close to trapping
frequency.\cite{ketterle2}\cite{cornell} It has been observed that
such rapid rotating condensate contains a large number of vortices
and they form a regular triangular lattice. The fast rotating BEC
is also characterized by two regimes. One is called the
Thomas-Fermi mean field (TFMF) regime, where the interaction
energy dominates over the kinetic energy. The other is the quantum
Hall mean field (QHMF) regime, in which the energy gap between
single particle Landau levels is much larger than the interaction
energy.\cite{Ho} Based on the quantum Hall mean field theory the
two-component fast rotating BEC was firstly studied by Mueller and
Ho.\cite{Mueller} The relative displacement $r_{0}$ between the
two sets of vortex lattices was found to be a non-vanishing value
for even a little positive inter-special interaction $U_{12}$.
With the increase of $U_{12}$ the vortex lattice will experience a
structure transition from triangular to square.\cite{Mueller}
Similar conclusions was later obtained by numerical study in the
Thomas-Fermi regime.\cite{ueda} All these have motivated us to
present a novel model that contains the above two physical
features and illustrates their interplay.

\section{The Model}

\begin{figure}[htbp]
\begin{center}
\includegraphics[width=1.2in]
{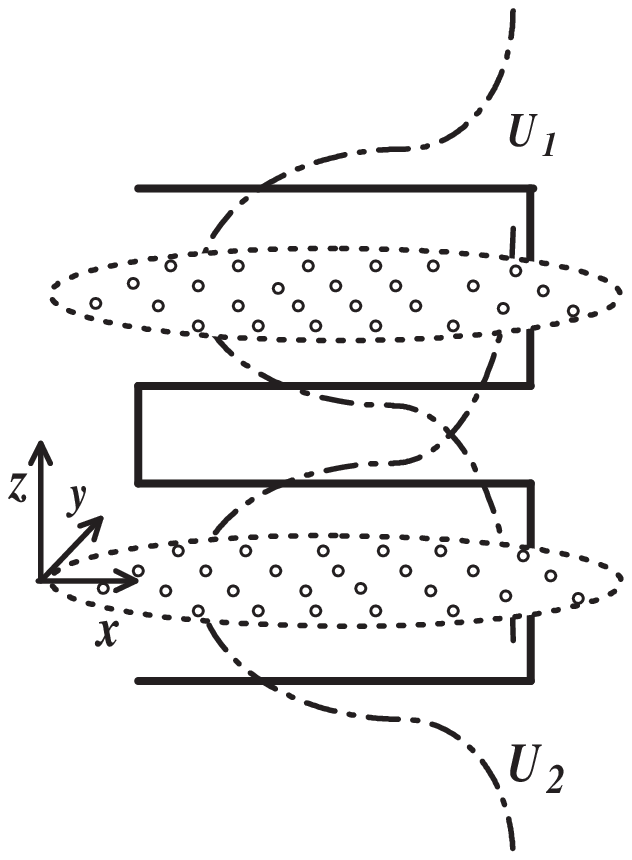}
\includegraphics[width=0.5in]
{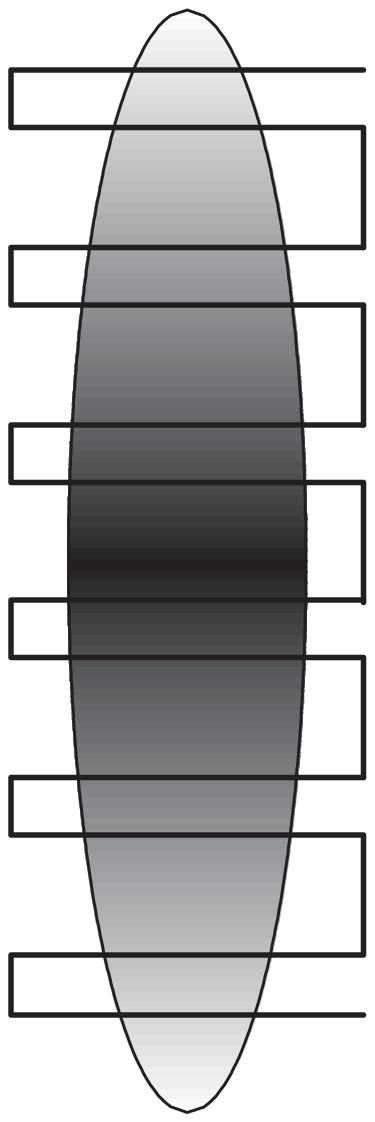} \caption{Left: The double-layer vortex lattice
system; \ \ Right: The rotating BEC confined in a optical lattice
in the $z$-direction.\label{illustrate}}
\end{center}
\end{figure}

In this section we introduce our model for double-layer rotating
BEC. The BEC is confined in a harmonic trap in the $xy$ plane and
high angular momentum with respect to $z$ direction is imparted
into the condensate. Along the $z$ direction the condensate is
confined in a double well potential. Thus two sets of vortex
lattices are formed in the wells and they couple to each other via
quantum tunnelling through the barrier between the wells. We call
such a system double-layer vortex lattice and schematically
illustrate it in Fig.\ref{illustrate}. It is believed to be
experimentally realizable due to the recent technique progress on
double well potential\cite{Ketterledwell}\cite{Kiltzing}. The
conclusions drawn from this model can be directly generalized to
the case that a fast rotating BEC is cut into pieces by applying
an optical lattice along the $z$-direction as shown in
Fig.\ref{illustrate}.

When these two layers are well separated by a high potential
barrier, the wave function overlapping along the $z$-direction is
sufficiently small, the phase fluctuation between the wells is
strong enough that the inter-layer phase coherence is lost and the
coherent hopping is suppressed. This regime is called Fock regime
because the particle number in each well is almost fixed. From the
work of Mueller and Ho \cite{Mueller} we know that in this regime
the relative displacement $r_{0}$ will take a non-zero value due
to the inter-layer density-density interaction. However this
effect may be frangible because this interaction is very weak in
this case.

While in the coherent regime the whole system is described by a
macroscopic wave function
\begin{equation}
\Psi=\sqrt{2}\left[\sqrt{n_{1}}u_{1}(z)\varphi_{1}(x,y)+\sqrt{n_{2}}e^{i\theta}u_{2}(z)\varphi_{2}(x,y)\right)]\label{macroscopoc
wave function}
\end{equation}
and the energy functional in the rotating frame is written as
\begin{equation}
H=\int dxdydz
\left[\Psi^{*}(h_{z}+h_{x,y}-\mu)\Psi+\frac{2\pi\hbar^2
a_{sc}N}{m}|\Psi|^4\right],\label{energy functional}
\end{equation}
where
\begin{equation}
h_{z}=-\frac{\hbar^2}{2m}\frac{d^2}{dz^2}+V_{\text{dou}}.
\end{equation}
Here $V_{\text{dou}}$ is the double well potential.
$u_{1,2}=(u_{s}\pm u_{a})/2$ represent the wave packets localized
in each well. Here $u_{s}$ and $u_{a}$ are the normalized ground
state and the first excited state of $h_{z}$, whose eigenvalues
are assumed to be far below other eigenvalues. $u_{s}$ and $u_{a}$
are symmetric and antisymmetric respectively, the energy splitting
between them is denoted by $t$.

In the rotating frame, the single particle Hamiltonian in $xy$
plane, $h_{x,y}$, is
\begin{equation}
h_{x,y}=-\frac{\hbar^2}{2m}\nabla^2+\frac{1}{2}m\omega_{\perp}
^2r^2-\Omega L_{z}
\end{equation}
where $\omega_{\perp}$ is the trapping frequency and $\Omega$ is
the rotation frequency. In the fast rotating condensate, $\Omega$
is quite close to $\omega_{\perp}$ and the condensate in each well
forms a vortex lattice, denoted by $\varphi_{1}$ and $\varphi_{2}$
respectively. We define $\Delta N=(n_{1}-n_{2})N$ as the relative
atom number difference between the two condensates, and $\theta$
is the relative phase. Corresponding to a normalized $\Psi$,
$n_{1}+n_{2}$ should be equal to unity.

Substituting the order parameter(\ref{macroscopoc wave function})
into the energy functional(\ref{energy functional}), the energy
functional can be written as follows:\cite{approximation}
\begin{widetext}
\begin{eqnarray}
E&=&-t\sqrt{n_{1}n_{2}}\left(\cos\theta\Re\int
dxdy\varphi_{1}^{*}\varphi_{2}+\sin\theta\Im\int
dxdy\varphi_{1}^{*}\varphi_{2}\right)+\int
dxdy\left[n_{1}\varphi_{1}^{*}h_{xy}\varphi_{1}+n_{2}\varphi_{2}^{*}h_{xy}\varphi_{2}\right]\nonumber
\\&+&4U\left(n_{1}^2\int dxdy|\varphi_{1}|^4+n_{2}^2\int
dxdy|\varphi_{2}|^4\right)\nonumber\\&+&8U_{12}n_{1}n_{2}\left(2\int
dxdy|\varphi_{1}|^2|\varphi_{2}|^2+\cos 2\theta\Re\int
dxdy(\varphi_{1}^{*}\varphi_{2})^2-\sin 2\theta\Im\int
dxdy(\varphi_{1}^{*}\varphi_{2})^2\right).\label{energy}
\end{eqnarray}
\end{widetext}
$\Re$ and $\Im$ stand for the real part and imaginary part
respectively. Here we denote $U=2\pi\hbar^2 a_{sc}N(\int
dzu_{1,2}^4)/m$ and $U_{12}=2\pi\hbar^2 a_{sc}N(\int
dzu_{1}^2u_{2}^2)/m$. In this case $U$ is always larger than
$U_{12}$ by at least one order of magnitude, so the energy minimum
occurs at $n_{1}=n_{2}$ and the two layers have the same lattice
type.

The main goal of this paper is to perform a systematic study of
the double-layer vortex lattice system based on the energy
functional(\ref{energy}), including the mean field phase diagram,
phase transition and excitations. Several key points will be
contained. Let us first briefly introduce them in the following.

As shown in the case of two-component rotating
BEC\cite{Mueller}\cite{ueda}, the repulsive density-density
interaction term $|\varphi_{1}|^2|\varphi_{2}|^2$ favors a
situation in which the high density region of one condensate
coincides with the low density region of the other, therefore the
vortices of the two condensates should avoid each other. The
difference between the current model and the two-component case is
the presence of coherent hopping terms such as
$\varphi_{1}^{*}\varphi_{2}$ and
$\varphi_{1}^{*2}\varphi_{2}^{2}$, which favor the vortices in two
layers being coincident. Hence the double-layer model is of
interest because there exist competing terms in the energy
functional, and it is natural to ask whether and how this
competition will manifest itself in a quantum phase transition.

We find out in this paper that such a transition does not always
take place, and we will answer the question under which condition
the competition will result in a phase transition. Because a
qualitative investigation of the inter-layer physics directly
relies on how we describe the vortex lattice state, and we know
that there exist two different mean field regimes for the fast
rotating BEC called Thomas-Fermi regime and quantum Hall regimes,
which are distinguished by the radio of the kinetic energy to the
interaction energy, we study the issue with two kinds of mean
field theories respectively and obtain two different criteria for
the inter-layer transition taking place in these two regimes. We
will remark in the end of this paper that the difference reveals
some intrinsic properties of these two mean field ansatzs.

The model studied here is also different from the double well
model because the condensate in each well has a vortex lattice
structure of its own, which is beyond the single model
approximation used in the discussion of double-well BEC. It is
known that the transition from coherent regime to Fock regime is
driven by the hopping element between two condensates. Here we
will find that the hopping element, as well as relative phase
fluctuation, is not only dependent on the wave function
overlapping along the $z$-direction as in the double well case,
but also depend on the integral $\int
dxdy\varphi_{1}^{*}\varphi_{2}$. Therefore, when two sets of
vortex lattices do not coincide with each other, the integral will
be relatively small and the tunnelling between two layers will be
suppressed. This presents a new mechanism for the transition from
coherent regime to Fock regime, which is another key point
discussed in this paper.

For the case that the transition can occur, we analytically give
the phase boundary between the coincident lattice phase and the
staggered lattice phase. For the case that the transition is
absent, we find that the relative phase $\theta$ between two
condensates will experience a second order change as $t/U_{12}$
decreases. After investigating a new excitation mode called vortex
lattice's sliding mode, we point out that such a change will
manifest itself in the frequency of sliding mode.

The paper is organized as follows. In the following section, we
will first study the case that the vortex lattice state is in the
QHMF regime. After a brief review of QHMF theory in the first
subsection \ref{QHMFA}, we focus on the transition for triangular
vortex lattice in the subsection \ref{QHMFB}, and we discuss the
sliding mode in the subsection \ref{QHMFC}. Then in the next
subsection \ref{QHMFD} we will generalize our discussion to
arbitrary lattice type and summarize the conclusions in the
subsection \ref{QHMFE}. In the section \ref{TFMF}, we focus on
TFMF regime and present a wave function ansatz to describe vortex
lattice state with the help of Thomas-Fermi approximation. Using
this ansatz we will revisit the issue studied in the third
section. The results obtained from the two different mean field
theories are compared. We remark in the last section that these
difference distinguish the intrinsic properties of the two
regimes.

\begin{figure}[htbp]
\begin{center}
\includegraphics[width=2.0in]
{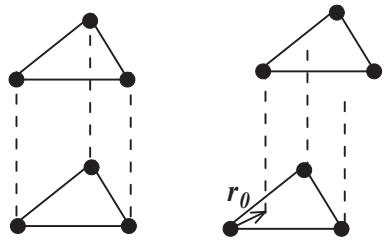} \caption{Left: The coincident lattice phase. Right: The
staggered lattice phase. The main part of this paper are concerned
with under which condition the transition between these two phases
can take place and determining the phase boundary.
\label{twophase}}
\end{center}
\end{figure}

\section{Quantum Hall Mean Field Regime}

\subsection{\label{QHMFA}Review of QHMF Theory}

Here we first briefly review the quantum Hall mean field theory
for single rotating condensate. We notice that $h_{x,y}$ can be
rewritten as
\begin{equation}
h_{x,y}=\frac{1}{2m}\left(-i\hbar\nabla-m\omega_{\perp}
\hat{z}\times \vec{r}\right)^2+\hbar(\omega_{\perp}-\Omega)L_{z} \
\ . \label{hxy}
\end{equation}
The Hamiltonian is identical to that describing a two-dimensional
particle moving in a perpendicular magnetic field in the symmetric
gauge. The eigenstates of $h_{x,y}$ are the Landau levels.
Defining the complex variable $w=(x+iy)/2a_{\perp}$ and
$w^{*}=(x-iy)/2a_{\perp}$ with
$a_{\perp}=\sqrt{{\hbar}/{m\omega_{\perp}}}$, in the quantum Hall
mean field picture the system is forced into the lowest Landau
level (LLL) under the condition that $\Omega$ is close to
$\omega_{\perp}$ and $\hbar\Omega$ is much larger than the
interaction energy, and consequently the macroscopic wave function
$\varphi_{1,2}(x,y)$ are analytical functions of the complex
variable $w$ apart from a Gaussian factor. In the LLL, the first
term in Eq.(\ref{hxy}) gives the same expectation value for all
states and can be neglected from the energy functional. The
expectation value of the single particle Hamiltonian turns out to
be\cite{Ho}
\begin{equation}
\langle h_{x,y}\rangle=\hbar(\omega_{\perp}-\Omega)\langle
|w|^2\rangle . \label{kinetic energy}
\end{equation}

It is noticed that for the vortex lattice states all first order
zeros of the entire functions $\varphi$, which are the locations
of vortices, form a regular lattice. Besides, the condensate
density should be circular symmetric if the confinement potential
is isotropic. These two requirements uniquely determine that
$\varphi$ should be in the following form:
\begin{eqnarray}
\varphi_{\text{qh}} = \theta(\zeta,\tau) \exp\left(\frac{\pi
w^2}{2v_{c}}\right) \exp\left(-\frac{r^2}{2a_{\perp}^2}\right)
 . \label{QHansatz}
\end{eqnarray}
Here $\theta(\zeta, \tau)$ is the Jacobi theta function. We denote
$b_{1}$ and $b_{2}$ as the basis vectors of the lattice,
$\tau=b_{2}/b_{1}=u+iv$ characterizes the lattice type and
$v_{c}=b_{1}^2v$ is the area of a unit cell. The argument $\zeta$
of the theta function is defined by scaling $\omega$ with $b_{1}$,
i.e. $\zeta=w/b_{1}$. The explicit form of $\theta(\zeta,\tau)$ is
\begin{equation}
\theta(\zeta,\tau)=\frac{1}{i}\sum\limits_{n=-\infty}^{+\infty}(-1)^ne^{i\pi\tau(n+\frac{1}{2})^2+2\pi
i\zeta (n+\frac{1}{2})} . \label{theta function}
\end{equation}

In Ref.\cite{Mueller} it has been shown that from Eq.(\ref{theta
function}) one can obtain that
\begin{equation}
|\varphi|^2=\frac{1}{\pi\sigma^2}\sum\limits_{\vec{K}}g_{K}\exp(i\vec{K}\cdot\vec{r})\exp(-r^2/\sigma^2),\label{density}
\end{equation}
where $\vec{K}_{m_{1}m_{2}}=m_{1}\vec{K}_{1}+m_{2}\vec{K_{2}}$ are
the reciprocal lattice vectors,
\begin{equation}
g_{K}=(-1)^{m_1+m_{2}+m_{1}m_{2}}e^{-v_{c}|K|^2/8\pi}\sqrt{v_{c}/2}\label{gK}
\end{equation}
and
\begin{equation}
v_{c}|K|^2=(2\pi)^2v^{-1}[(vm_{1})^2+(m_{2}+um_{1})^2].
\end{equation}
The condensate radius $\langle r^2\rangle$ is modified to
$\sigma^2$ by the presence of large number of vortices, and
\begin{equation}
\sigma^{-2}=a_{\perp}^{-2}-\pi v_{c}^{-1}.
\end{equation}

Furthermore, with the help of Eq.(\ref{density}) we can evaluate
the self-interaction energy of each condensate
\begin{equation}
\int
dxdy|\varphi|^4=\frac{1}{2\pi\sigma^2}\frac{\sum\limits_{K,K^\prime}g_{K}g_{K^\prime}e^{-\sigma^2|K+K^\prime|^2/2}}
{\left(\sum\limits_{K}g_{K}e^{-\sigma^2{K}^2/4}\right)^2} .
\label{self-interaction}
\end{equation}
In the larger vortex number limit, $\pi\sigma^2/v_{c}\gg 1$, we
only keep the $K=-K^\prime$ terms in the numerator and $K=0$ term
in the denominator. The integral(\ref{self-interaction}) can then
be
 simplified to
\begin{equation}
\int
dxdy|\varphi|^4=\frac{1}{2\pi\sigma^2}\sum\limits_{K}\left|\frac{g_{K}}{g_{0}}\right|^2
=\frac{I_{1}}{2\pi\sigma^2}.\label{I}
\end{equation}

Then we can minimize the energy functional, including
Eq.(\ref{kinetic energy}) and Eq.(\ref{I}), to obtain the average
vortex density and the vortex lattice structure. We notice that
all the information about the lattice type is contained in
$I_{1}$. Minimizing $I_{1}$ we will find the lattice structure
being triangular. However, if some additional factors are
included, such as the anisotropy of the confinement potential, it
is also possible for the lattice structure changing from
triangular to others.\cite{stript}\cite{StriptEx}\cite{MOktel}
%also notice that $g_{K}$ will be exponential small as $K$ increasing,
%the dominator contribution to the summation comes from $K_{00}$,
%$K_{0,\pm 1}$, $K_{\pm 1,0}$ and $K_{\pm 1,\pm 1}$, we can

\subsection{\label{QHMFB}Double-Layer Triangular Vortex Lattice}
Now we begin to discuss two coupled rotating BECs by using QHMF
theory. The inter-layer interaction and the coherent hopping
between two condensates should be considered. Because in this
model $U_{12}\ll U$ and the hopping terms are almost independent
of the lattice structure as we will show later, we can assume that
the inter-layer coupling will not change the lattice structure of
each condensate. In this subsection we will firstly focus on the
triangular lattice.

It has been found in Ref.\cite{Mueller} that
\begin{equation}
\int
dxdy|\varphi_{1}|^2|\varphi_{2}|^2=\frac{1}{2\pi\sigma^2}\sum\limits_{K}\left|\frac{g_{K}}{g_{0}}\right|^2
\cos\vec{K}\cdot\vec{r}_{0}.\label{inter-layer inter}
\end{equation}
Because $g_{K}$ exponentially depend on $|K|$ as one can see from
Eq.(\ref{gK}), only several terms such as $g_{00}$, $g_{0,\pm 1}$,
$g_{\pm,0}$ and $g_{\pm 1,\pm 1}$ need to be considered. For
triangular lattice we have $I_{1}=1.1596$, and
\begin{equation}
\int dxdy
|\varphi_{1}|^2|\varphi_{2}|^2=\frac{1}{2\pi\sigma^2}[1+f(\vec{r}_{0})],
\end{equation}
where
\begin{eqnarray}
f(\vec{r_{0}})=C\sum\limits_{K_{01},K_{10},K_{1-1}}\cos\vec{K}\cdot\vec{r}_{0}\label{f(r)}
\end{eqnarray}
with $C=0.0532$. The minimum value of $f(r_{0})$ is $-3C/2$ which
occurs at $r_{0}=(b_{1}+b_{2})/3$, and its maximum is $3C$
occurring at $r_{0}=0$.

Using Eq.(\ref{theta function}) we evaluate the coherence terms
(for details see the appendix \ref{A1}). Considering the number of
vortices contained in the cloud $N_{v}=\pi\sigma^2/v_{c}\gg 1$ and
following the same approximation made in Ref.\cite{Mueller}, we
drop all the $K\neq 0$ terms and obtain the simplified expression
of Eq.(\ref{hopping term}) is
\begin{equation}
\Re\int
dxdy\varphi^{*}_{1}\varphi_{2}=\exp\left(-N_{v}\frac{\pi}{v_{c}}r_{0}^2\right)
\end{equation}
and
\begin{equation}
\Im\int dxdy\varphi^{*}_{1}\varphi_{2}=0.
\end{equation}
Similarly we obtain the real part of the two-particle coherent
hopping term as
\begin{eqnarray}
\Re\int dxdy (\varphi_{1}^{*}\varphi_{2})^2&=&
\frac{1}{2\pi\sigma^2}\sum\limits_{K}\left|\frac{g_{K}}{g_{0}}\right|^2
\exp\left(-N_{v}\frac{\pi}{2v_{c}}r_{0}^2\right)\nonumber\\
&=&\frac{I_{1}}{2\pi\sigma^2}\exp\left(-N_{v}\frac{\pi}{2v_{c}}r_{0}^2\right)
\end{eqnarray}
while its imaginary part vanishes. These coherent terms are
exponentially dependent on the relative displacement $r_{0}$, and
the exponential factor $N_{v}$ is very large for the case
discussed in this paper. This indicates that the coherent terms
will be exponentially small and the phase fluctuation will be
enhanced when $r_{0}^2$ is comparable to $v_{c}$. It means that if
the competition results in a transition from $\vec{r}_{0}=0$ to a
non-vanishing $\vec{r}_{0}$, such a transition will be accompanied
by the loss of phase coherence.

According to above discussions, the energy functional can be
rewritten as following:
%\begin{widetext}
\begin{eqnarray}
&&E=-\frac{t}{2}\cos\theta
\exp\left(-N_{v}\frac{\pi}{v_{c}}r_{0}^2\right)
+\hbar(\omega-\Omega)\frac{\sigma^2}{a_{\perp}^2}\nonumber\\&&+\frac{1}{2\pi\sigma^2}
[2UI+4U_{12}+4U_{12} f(r_{0})\nonumber\\&&+2U_{12}I\cos 2\theta
\exp\left(-N_{v}\frac{\pi}{2v_{c}}r_{0}^2\right)].\nonumber\\\label{energy
function}
\end{eqnarray}
%\end{widetext}

In the following we will minimize the energy functional to obtain
the mean field ground state, and then discuss the phase diagram.
First of all we minimize $E$ with respect to $\sigma$. As
$U_{12}/U$ is a small parameter, we can expand the result to its
first order, which yields that
\begin{equation}
E=-\frac{t}{2}\cos\theta e^{-N_{v}\frac{\pi
r_{0}^2}{v_{c}}}+\frac{AU_{12}}{\sqrt{I_{1}}U}\left[f(r_{0})+\frac{I_{1}}{2}\cos2\theta
e^{-N_{v}\frac{\pi r_{0}^2}{2v_{c}}}\right].
\end{equation}
Here $A$ is defined as $\sqrt{U(\omega_{\perp}-\Omega)/(\pi
a_{\perp}^2)}$. Two constant terms, $2A\sqrt{I_{1}}$ and
$AU_{12}/(\sqrt{I_{1}}U)$ have been neglected in the above
expression.

When
%\begin{equation}
$t>4A\sqrt{I_{1}}U_{12}e^{-N_{v}\frac{\pi r_{0}^2}{2v_{c}}}/U$,
%\end{equation}
the expectation value of the relative phase $\bar{\theta}$ tends
to be zero, and then
\begin{equation}
E=-\frac{t}{2}\left(e^{-N_{v}\frac{\pi
r_{0}^2}{2v_{c}}}-\frac{AU_{12}\sqrt{I_{1}}}{2Ut}\right)^2+\frac{A^2U_{12}^2I_{1}}{8U^2t}
+\frac{AU_{12}}{\sqrt{I_{1}}U}f(r_{0}). \label{Er1}
\end{equation}
While when $t<4A\sqrt{I_{1}}U_{12}e^{-N_{v}\frac{\pi
r_{0}^2}{2v_{c}}}/U$, $\bar{\theta}$ will gradually change from
zero to $\pi/2$ with $t$ decreasing, and
\begin{equation}
\cos\bar{\theta}=\frac{tU}{4\sqrt{I_{1}}AU_{12}}\exp\left(-N_{v}\frac{\pi}{2v_{c}}r_{0}^2\right),
\end{equation}
the energy minimum will be
\begin{equation}
E=-\frac{t^2Ue^{-N_{v}\frac{3\pi
r_{0}^2}{2v_{c}}}}{16AU_{12}\sqrt{I_{1}}}-\frac{AU_{12}\sqrt{I_{1}}}{2U}e^{-N_{v}\frac{\pi
r_{0}^2}{2v_{c}}}+\frac{AU_{12}}{\sqrt{I_{1}}U}f(r_{0}).\label{Ertw}
\end{equation}

\begin{figure}[htbp]
\begin{center}
\includegraphics[width=2.3in]
{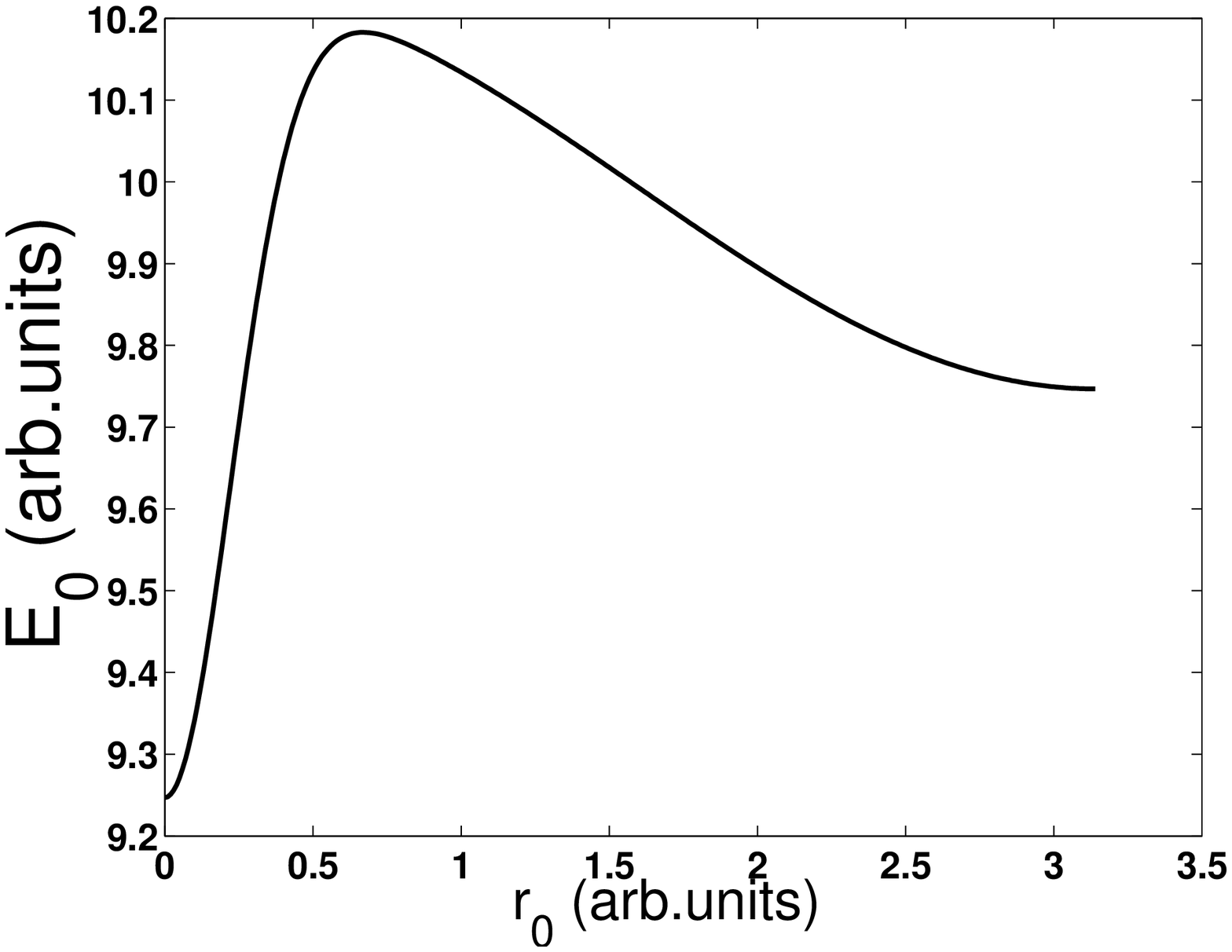}
\includegraphics[width=2.3in]
{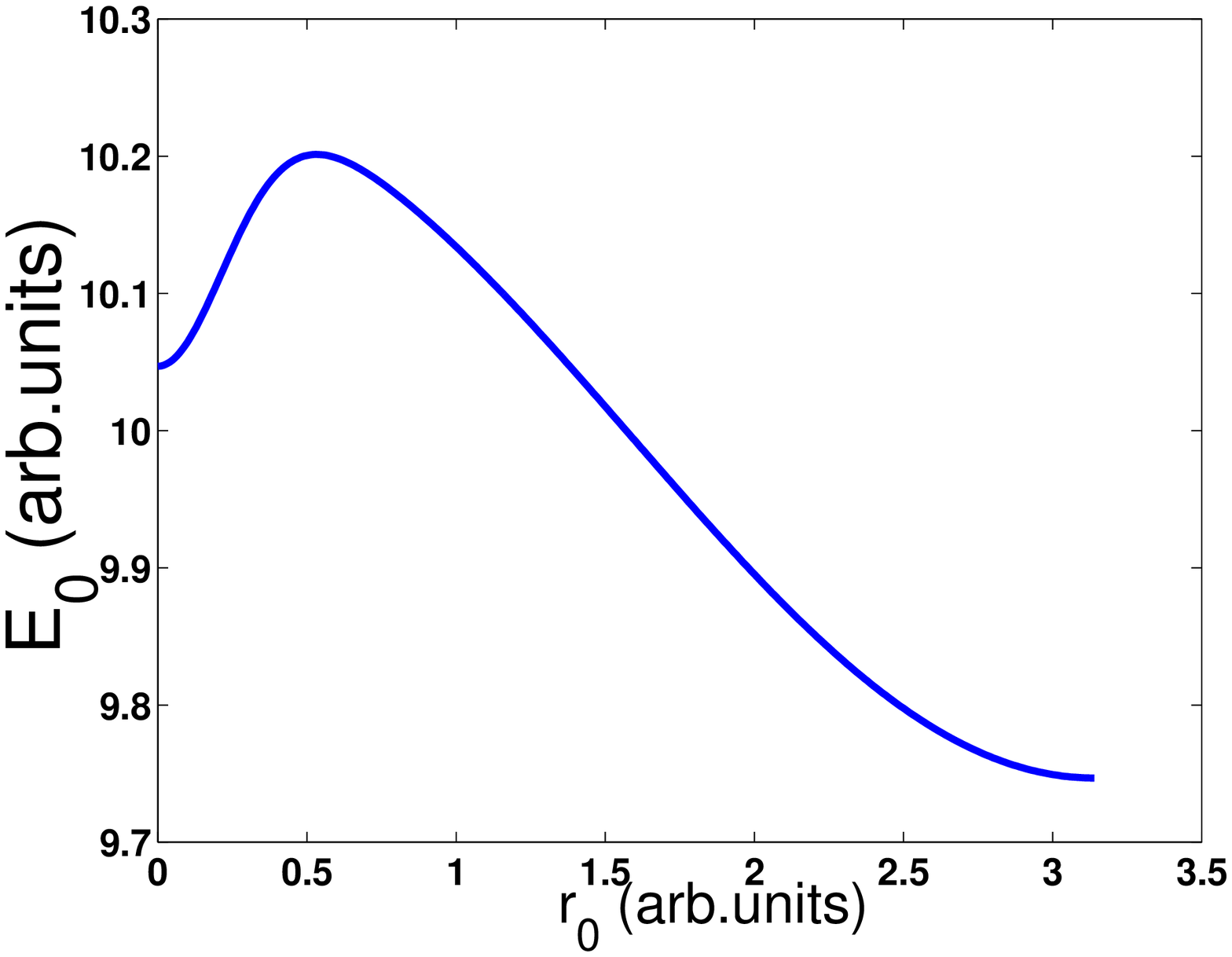}\caption{The minimum value of the energy functional
$E$ as a function of $r_{0}$, Eq.(\ref{Er1}) and Eq.(\ref{Ertw}),
has two possible shapes. In the large $t$ limit it behaves like
the upper one, and one may expect that it will eventually behave
like the lower one as $t/U_{12}$ decreasing. In the following text
we will discuss whether and when such a transition will take
place. \label{schematically}}
\end{center}
\end{figure}

From the $r_{0}$ dependence of the energy Eq.(\ref{Er1}) and
Eq.(\ref{Ertw}), one can easily find that the hopping energy favor
a vanishing $r_{0}$ while $f(r_{0})$ favors a non-vanishing
$r_{0}$. As schematically shown in Fig.(\ref{schematically}) there
possibly exist two local minima located at $r_{0}=0$ and
$r_{0}=(b_{1}+b_{2})/3$. The true ground state is determined by
comparing the two local minima, denoted by $E_{1}$ and $E_{2}$
respectively. In the large $t$ limit when the hopping energy is
dominant, the ground state is at $\vec{r}_{0}=0$ as shown in the
left side of Fig.(\ref{schematically}). One may expect that
$E_{1}$ will be eventually larger than $E_{2}$ as the right side
one, and result in a first order transition of $r_{0}$ as
$t/U_{12}$ decreasing.

At $r_{0}=(b_{1}+b_{2})/3$ the coherent terms are small enough and
can be neglected, the minimum value of $E$ is therefore
independent of $t$ and $\theta$. One can easily find out from
Eq.(\ref{Er1}) and Eq.(\ref{Ertw}) that
\begin{equation}
E_{2}=-\frac{3C}{2}\frac{AU_{12}}{\sqrt{I_{1}}U}.
\end{equation}
We can also find from these two equations that $E_{1}$ increases
as $t$ decreases, however there exists an upper bound, i.e.
\begin{equation}
E_{1}\leq
\frac{AU_{12}}{\sqrt{I_{1}}U}3C-\frac{AU_{12}}{2U}\sqrt{I_{1}}.
\end{equation}
For the phase transition taking place, it is required that
$E_{1}>E_{2}$, i.e
\begin{equation}
9C>I_{1} \label{transition condition}.
\end{equation}
This condition can not be satisfied for the triangular vortex
lattice, where $I_{1}=1.1596$ and $C=0.0532$. Therefore, the
conclusion is that the two sets of vortex lattices will coincide
throughout all of the coherent regime.

As we have emphasized the underlying physics of the transition of
$r_{0}$ is the competition between the coherent terms and the
density-density interaction term. The coherent terms include
one-particle hopping term $\varphi_{1}^*\varphi_{2}$, whose
characteristic energy is $t$, and the two-particle hopping terms
$\varphi_{1}^{* 2}\varphi_{2}^2$, whose characteristic energy is
$U_{12}$. When $t\gg U_{12}$ the relative phase $\theta$ between
two layers will be zero, which is beneficial to one-particle
hopping energy. We found that in this case the hopping energy will
always dominate over the density-density interaction energy. With
$t$ decreasing, $\theta$ will change from zero to $\pi/2$ to
minimize the two-particle hopping energy. Eventually whether the
transition can occur mostly depends on the competition between
two-particle hopping term and the density-density interaction
term. Since both these two terms share the same characteristic
energy $U_{12}$, which term will be dominant relates to the
intrinsic properties of the vortex lattice states. This can be
seen from Eq.(\ref{transition condition}), where $I$ and $C$ only
depend on the lattice structure.

Hence if the lattice type is changed, the above conclusion must
consequently be changed. We will generalize the above discussion
to arbitrary lattice type and find out on what condition the
transition can occur in the fourth subsection.

\subsection{\label{QHMFC}Sliding Mode Frequency}
Before continuing our discussion on the transition, in this
subsection we discuss briefly one new excitation mode, which is
unique to this system. This mode is characterized by small
amplitude oscillation of $r_{0}$ around its equilibrium position
$r_{0}=0$. We call this mode sliding mode because the vortices in
one layer oscillate in locked steps relative to the other layer,
and for the optical lattice case (see the right side of
Fig.(\ref{illustrate})) this mode can be corresponding to Kelvin
mode of vortex line in the conventional three-dimensional rotating
BEC.\cite{Stoof}

We write down the propagate in the path integral representation as
follows
\begin{equation}
Z=\int d\Psi^{*}d\Psi \exp\left[\frac{i}{\hbar}\int dt d^2r
i\Psi^*\partial_{t}\Psi-E(\Psi^*,\Psi)\right].\label{Lagrangian}
\end{equation}
To consider the oscillation of $r_{0}$ we have
\begin{equation}
\Psi^*\partial_{t}\Psi=\Psi^*(\partial_{\delta
r_{0}}\Psi)\delta\dot{r}_{0}
=n_{2}\varphi_{2}^*(\partial_{r_{0}}\varphi_{2})\delta\dot{r}_{0},
\end{equation}
the second equality follows from the fact that only $\varphi_{2}$
contains $r_{0}$. We notice that it is necessary to go beyond the
mean field approximation and take the density fluctuation into
account in order to obtain a dynamic term of $r_{0}$. The
fluctuation of $n_{2}$ denoted by $\delta n$ contains two
contributions. One comes from the inter-layer coupling, which is
independent of spatial coordinates. The coupling to this part can
not induce a dynamic term of $r_{0}$ because $\int d^2r
\varphi^*\partial_{r_{0}}\varphi$ vanishes at $r_{0}=0$. Therefore
in order to obtain a non-vanishing dynamic term we should include
the intra-layer local density fluctuation bringing the condensate
out of the lowest Landau levels. The dynamic term in
Eq.(\ref{Lagrangian}) is then written as
\begin{equation}
\int dt \delta\dot{r}_{0}\int d^2r
\delta{n}(r)i\varphi_{2}^*\partial_{r_{0}}\varphi_{2}.
\end{equation}

We can then expand the energy functional around the saddle point
to the second order, the result is
\begin{eqnarray}
\delta^2E&=&\frac{t}{2}\cos\bar\theta N_{v}\frac{\pi}{v_{c}}\delta
r_{0}^2-\frac{U}{\pi\sigma^2_{0}}I_{1}\cos2\bar\theta
N_{v}\frac{\pi}{2v_{c}}\delta r_{0}^2\nonumber\\
&+&\int d^2r4U|\varphi_{2}|^4\delta n(r)^2.
\end{eqnarray}
The term $f(r_{0})$ has been neglected in the above equation
because the coefficient of second order expansion in the
neighborhood of $r_{0}=0$ is relative small. Here $\sigma_{0}$ is
the saddle point of $\sigma$. We first integrate the field $\delta
n(r)$ out to obtain a dynamic term of $r_{0}$, the effective
Lagrangian describing the oscillation of $r_{0}$ then reads
\begin{equation}
L=\int
\frac{m_{\text{eff}}}{2}\delta\dot{r}_{0}^2-\frac{N_{v}\pi}{2v_{c}}\left[t\cos\bar\theta
-\frac{UI_{1}\cos2\bar\theta }{\pi\sigma^2_{0}}\right]\delta
r_{0}^2.
\end{equation}
Here $m_{\text{eff}}$ is the effective dynamic mass of the
collectively oscillation motion defined by
\begin{equation}
m_{\text{eff}}=\int
d^2r\frac{-(\varphi_{2}^*\partial_{r_{0}}\varphi_{2})^2}{8U|\varphi_{2}|^4}.
\end{equation}
Hence the oscillation frequency is
\begin{equation}
\omega_{s}=\sqrt{\frac{N_{v}\pi}{v_{c}}\left[t\cos\bar\theta
-\frac{UI_{1}\cos2\bar\theta
}{\pi\sigma^2_{0}}\right]\frac{1}{m_{\text{eff}}}}.
\end{equation}

What should be emphasized here is that this frequency is a smooth
function of $\cos\bar\theta$. As shown in the above section
$\bar{\theta}$ will gradually change from zero to $\pi/2$ below
the critical point $t/U_{12}=4A\sqrt{I_{1}}/U$. This leads to the
key prediction of this subsection that the sliding mode frequency
will consequently exhibit a second order discontinuity at the
critical point.

\subsection{\label{QHMFD}General Lattice Type}
To discuss a vortex lattice with arbitrary lattice structure,
instead of Eq.(\ref{f(r)}) we define
\begin{equation}
f(r_{0})=2\sum\limits_{K_{01},K_{10},K_{1-1}}\left(\frac{g_{K}}{g_{0}}\right)^2\cos\vec{K}\cdot\vec{r_{0}},
\end{equation}
and denote its minimum by $M$, the condition(\ref{transition
condition}) for a transition in $r_{0}$ is then modified to
\begin{equation}
I_{1}-2-2M>0\label{transion condition general}.
\end{equation}
For flat lattice $g_{K_{1,-1}}$ is much larger than $g_{K_{0,1}}$
and $g_{K_{1,0}}$. As an example for the lattice with
$\arg\tau=\pi/6$,
\begin{equation}
\left|\frac{g_{K_{1,-1}}}{g_{0}}\right|^2=\left|\frac{g_{K_{-1,1}}}{g_{0}}\right|^2=0.1857,
\end{equation}
it is much larger than $|g_{K_{0,1}}/g_{0}|^2$ and
$|g_{K_{1,0}}/g_{0}|^2$, which are both equal to $0.0019$. In this
case the minimum of $f(r_{0})$ is approximately
\begin{equation}
M=-f(r_{0}=0)=1-I_{1},
\end{equation}
which can be reached when $r_{0}=(b_{1}+b_{2})/4$.

\begin{figure}[htbp]
\begin{center}
\includegraphics[width=2.5in]
{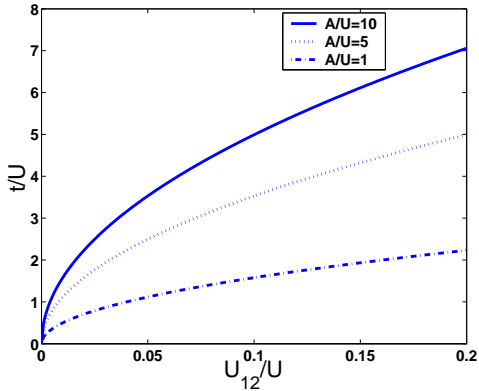} \caption{ The phase boundary between
$\vec{r}_{0}=0$ coherence regime and $\vec{r}_{0}\neq 0$ Fock
regime, for different values of $A/U$. In the regime above the
line $\vec{r}_{0}=0$ and below the line $\vec{r}_{0}\neq 0$, there
is a first order transition when crossing the line.
\label{transitionline}}
\end{center}
\end{figure}

Hence the condition Eq.(\ref{transion condition general}) can be
satisfied when the lattice is flat enough that satisfies
$\arg\tau<0.6728$. The transition takes place when $E_{1}=E_{2}$,
i.e.
\begin{equation}
\left(\frac{t}{U}\right)^2=\frac{U_{12}}{U}8(3I_{1}-4)\left(\frac{A}{U}\right)^2.
\end{equation}
Here $A/U$ is a dimensionless parameter and can be tuned in a wide
range. As an example, we show in Fig.(\ref{transitionline}) the
transition line for a flat lattice with $\arg\tau=\pi/6$. In the
upper side the vortices in two layers coincide with each other and
the vortex lines are parallel to the $z$ axis. On the other side
two sets of lattices are staggered, and the phase coherence
between layers is lost.

Therefore to experimentally observe such a transition in QHMF
regime, the first step is to produce a sufficient flat vortex
lattice. Although so far most observed vortex lattices are
triangular, it is still possible to achieve other lattice
structure by using some special methods. Recently in
Ref.\cite{MOktel} the author shown within the LLL approximation
that for a condensate with small number of vortices the lattice
structure can be compressed to be quite flat due to the anisotropy
of confinement potential.

\begin{figure}[htbp]
\begin{center}
\includegraphics[width=3.0in]
{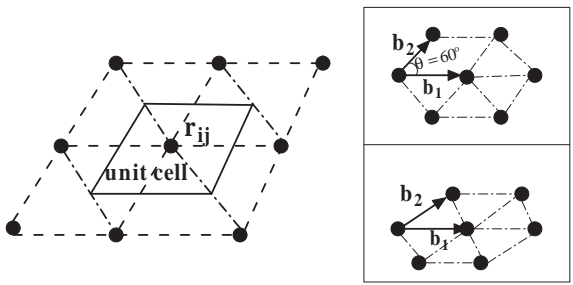} \caption{Left: The unit cell. Right: The schematic
exhibition of triangular lattice and flat
lattice.\label{latticetype}}
\end{center}
\end{figure}

\subsection{\label{QHMFE}Summarizes}

\begin{figure}[htbp]
\begin{center}
\includegraphics[width=2.3in]
{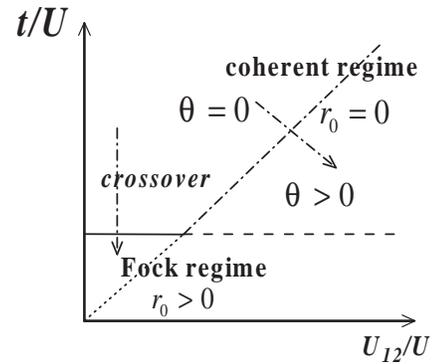}
\includegraphics[width=2.3in]
{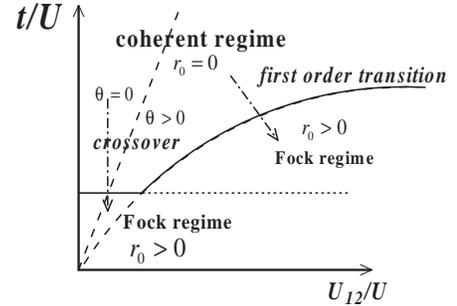} \caption{The phase diagram for the double-layer
vortex lattice. The upper is for triangular lattice and the lower
is for flat lattice \label{phasediagram}}
\end{center}
\end{figure}

In this section we have discussed the phase diagram of
double-layer vortex lattice state, when both rotating condensates
are in the QHMF regime. We summarize our results as follows and
the phase diagram is shown in Fig.(\ref{phasediagram}).

$1.$ In the coherent regime, for triangular lattice a transition
of $\vec{r}_{0}$ is absent and $\vec{r}_{0}$ will always remain
zero when the coherent hopping terms depending on the relative
phase between two layers are introduced into the energy
functional. When $t$ decreases the relative phase between two
condensates will experience a second order change. The change of
phase manifests itself in a second order change of the sliding
mode frequency. The phase diagram is shown in the left side of
Fig.(\ref{phasediagram})

$2.$ When the lattice is compressed to be quite flat, a transition
of $r_{0}$ will occur when $U_{12}/t$ exceeds some critical value.
The number fluctuation will be suppressed after the jumping of
$r_{0}$ and a first order transition from the coherent regime to
the Fock regime will be induced. It is noticed that such a
transition from coherent regime to Fock regime is different from
that discussed in the double-well BEC case, because it is driven
by $t/U_{12}$ instead of $t/U$, and it is a first order transition
instead of crossover. The phase diagram is shown in the right-side
of Fig.(\ref{phasediagram})

\begin{figure}[htbp]
\begin{center}
\includegraphics[width=2.5in]
{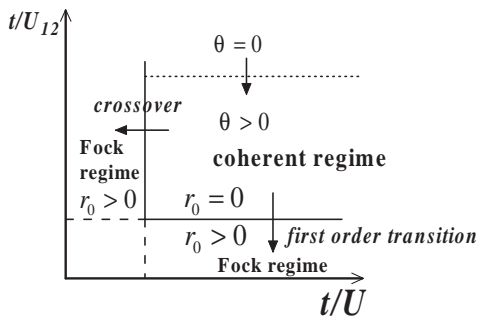} \caption{The phase diagram for the double-layer
vortex lattice. This phase diagram is obtain when each rotating
BEC is in Thomas-Fermi regime and the condition
Eq.(\ref{trancondiTF}) is satisfied. \label{phasediagram3}}
\end{center}
\end{figure}

\section{\label{TFMF}Thomas-Fermi Regime}

We have mentioned that there are two different mean field
descriptions of rotating BEC. In the last section we have focused
on the QHMF regime. Before investigating the same issue when both
condensates are in TFMF regime, we would like to have a brief
discussion on the difference between two regimes.

As shown in the last section, QHMF works when the kinetic energy
is much larger than the interaction energy. The wave functions are
eigenstates of single particle Hamiltonian, and the mean value of
$h_{xy}$ only depends on the average vortex density, and $\langle
h_{x,y}\rangle$ neither depends on the winding number of each
vortex nor the structure of vortex lattice. That the singly
quantized vortices are arranged into triangular lattice is
determined by taking the self-interaction energy into account. In
contrast, TFMF is obtained by neglecting the kinetic energy, and
the ground state density distribution is the result of the balance
between effective trapping energy and interaction energy. The
density distribution is assumed to be not remarkably changed and
keeps a form of an inverted parabola when large number of vortices
are contained. In this regime the interaction energy is
independent of both the vortex winding number and the lattice
structure. The kinetic energy caused by superfluid currents around
the centers of vortices will induce an effective repulsive
interaction between them. As a result the vortices are singly
quantized and arrange into triangular lattice.

Besides there are also phenomenological differences between two
regimes. In the TFMF regime the density profile is an inverted
parabola and there exists a vortex core structure for each vortex.
A vortex core structure means that there is a characteristic
length $\xi$, inside which the superfluid density drops to zero
rapidly, and outside which the superfluid density recovers the
value free from vortices. Because the healing length increases as
interaction strength decreases, when entering the QHMF regime
$\xi$ is comparable to the inter-vortex spacing and no longer well
defined. The individual vortex core structure disappears.

Therefore to deal with the TFMF regime we should use another
wavefunction ansatz, which will be quite different from the one in
QHMF regime. We know that the Thomas-Fermi density profile is
\begin{equation}
\rho_{0}(r)=\frac{1}{2g}\left(\mu-\frac{1}{2}m(\omega^2-\Omega^2)r^2\right).
\end{equation}
In the presence of vortex lattice we assume that the density
profile is modified in the following way
\begin{equation}
\rho\propto
\rho_{0}-\sum\limits_{r_{ij}}\rho_{o}(r_{ij})\exp\left(-\frac{(r-r_{ij})^2}{\xi^2}\right),
\end{equation}
where $r_{ij}$ is the centers of vortices. $\xi$ is the vortex
core side. Here we assume the $\xi$ is the same for all vortices.
This ansatz is believed to be valid in the regime $\pi\xi^2<
v_{c}\ll \pi R^2$, where $R$ is the Thomas-Fermi radius of the
condensate.

We notice that
\begin{eqnarray}
&&\int
d^2\vec{r}\left[\rho_{0}-\sum\limits_{r_{ij}}\rho_{o}(r_{ij})\exp\left(-\frac{(r-r_{ij})^2}{\xi^2}\right)\right]\nonumber\\
&&=1-\sum\limits_{r_{ij}}\rho_{0}(r_{ij})\pi\xi^2,\label{normalization}
\end{eqnarray}
in the large vortices limits, we can assume the area of a unit
cell is much smaller than the size of condensate, approximately we
have
\begin{equation}
\sum\limits_{r_{ij}}\rho_{0}(\vec{r}_{ij})v_{c}=\int
d^2\vec{r}\rho_{0}(\vec{r})=1.
\end{equation}
Hence Eq.(\ref{normalization}) reduces to $1-\pi\xi^2/v_{c}$. Such
approximation will be often used in the following derivation, the
spirit of which is essentially keeping accuracy to the first order
of $\pi\xi^2/v_{c}$ and neglecting $v_{c}/(\pi R^2)$. Similar
approach has also been used in Ref.\cite{Baym} in the study of
single component vortex lattice state.

The density distribution of vortex lattice state is
\begin{equation}
\rho=\frac{1}{1-\frac{\pi\xi^2}{v_{c}}}
\left[\rho_{0}-\sum\limits_{r_{ij}}\rho_{o}(r_{ij})\exp\left(-\frac{(r-r_{ij})^2}{\xi^2}\right)\right],
\end{equation}
and the wave function is written as
\begin{equation}
\varphi=\sqrt{\rho}\frac{f(z)}{|f(z)|}=\sqrt{\rho\frac{f(z)}{f(\bar{z})}}.
\end{equation}
Here $f(z)=\prod\limits_{z_{ij}}(z-z_{ij})$ and can be described
by the Jacobi theta function mentioned in the previous section.

In this regime, we can neglect the overlapping between vortex
cores and then have
\begin{eqnarray}
&&|\varphi_{1}(r)|^2|\varphi_{2}(r)|^2=\frac{1}{\left(1-\frac{\pi\xi^2}{v_{c}}\right)^2}\{\rho_{0}^2-
\nonumber\\&&\sum\limits_{r_{ij}} \rho_{0}(r)
\rho_{0}(r_{ij})e^{-(r-r_{ij})^2/\xi^2}\nonumber\\&&+\rho(r)\rho_{0}(r_{ij}+r_{0})e^{-(r-r_{ij}-r_{0})^2/\xi^2}
\nonumber\\&&
 +\rho_{0}(r_{ij})\rho_{0}(r_{ij}+r_{0})
e^{-[(r-r_{ij})^2+(r-r_{ij}-r_{0})^2]/\xi^2}\}.\label{TFdensity-density}
\nonumber\\
\end{eqnarray}
Approximately we can assume
$\rho_{0}(r_{ij}+r_{0})\simeq\rho_{0}(r_{ij})$ because $r_{0}$ is
less than one lattice spacing and much smaller than the condensate
size. Hence in Eq.(\ref{TFdensity-density}) only the last term
involves $r_{0}$, explicitly
\begin{eqnarray}
&&\int
dxdy\sum\limits_{r_{ij}}\rho^2(r_{ij})e^{-[(r-r_{ij})^2+(r-r_{ij}-r_{0})^2]/\xi^2}\nonumber\\
&=&\sum\limits_{r_{ij}}\rho^2(r_{ij})\frac{\pi\xi^2}{2}e^{-r_{0}^2/(2\xi^2)}\nonumber\\
&\simeq &
\frac{\pi\xi^2}{2v_{c}}I_{2}\exp\left(-\frac{1}{2}\frac{r_{0}^2}{\xi^2}\right),
\end{eqnarray}
where $I_{2}$ is defined as $\int \rho_{0}^2 d^2r$. And therefore
\begin{eqnarray}
\int
dxdy|\varphi_{1}|^2|\varphi_{2}|^2=\frac{1-2\frac{\pi\xi^2}{v_{c}}+\frac{\pi\xi^2}
{2v_{c}}e^{-\frac{1}{2}\frac{r_{0}^2}{\xi^2}}}{\left(1-\frac{\pi\xi^2}{v_{c}}\right)^2}I_{2}.
\end{eqnarray}
We notice that the larger $r_{0}^2$, the smaller the
density-density interaction energy. Hence its minimum is reached
when $r_{0}=(b_{1}+b_{2})/2$.

Now we begin to discuss the coherent hopping terms in TF regime
where
\begin{equation}
\varphi_{1}^{*}\varphi_{2}=\sqrt{\rho_{1}\rho_{2}}\sqrt{\frac{\varphi_{\text{qh}1}^{*}
\varphi_{\text{qh}2}}{\bar{\varphi}_{\text{qh}{1}}\bar{\varphi}_{\text{qh}2}^{*}}}\
\ ,
\end{equation}
using the result of Eq.(\ref{coherenthopping}) and taking the
approximation of dropping all the terms with $K\neq 0$, we obtain
that
\begin{equation}
\Re
\varphi_{1}^{*}\varphi_{2}=\sqrt{\rho_{1}\rho_{2}}\cos\left(\frac{\pi}{v_{c}}\vec{r}\times\vec{r}_{0}\right),
\label{hopterm}
\end{equation}
and
\begin{equation}
\Im
\varphi_{1}^{*}\varphi_{2}=\sqrt{\rho_{1}\rho_{2}}\sin\left(\frac{\pi}{v_{c}}\vec{r}\times\vec{r}_{0}\right).
\end{equation}
From a careful analysis in the appendix(\ref{A3}) we conclude that
at $r_{0}$=0
\begin{equation}
\Re \int dxdy \varphi_{1}^{*}\varphi_{2}=1,
\end{equation}
and
\begin{equation}
\Re \int dxdy \varphi_{1}^{*
2}\varphi_{2}^2=I_{2}\frac{1-\frac{3\pi\xi^2}{2v_{c}}}{(1-\frac{\pi\xi^2}{v_{c}})^2}
=I_{3}.
\end{equation}
Their imaginary parts vanish due to the spatial reflection
symmetry. These coherent terms decrease as $r_{0}$ increases. When
$r_{0}$ reaches its maximum value $(b_{1}+b_{2})/2$ these terms
are small enough that can be neglected.

In the following we will find the ground state by comparing the
minimum values of the energy function at $r_{0}=0$ and
$r_{0}=(b_{1}+b_{2})/2$. We notice that at $r_{0}=0$
\begin{equation}
E= -\frac{t}{2}\cos\theta+4U_{12}I_{3}+2U_{12}\cos2\theta I_{3}
\end{equation}
and at $r_{0}=(b_{1}+b_{2})/2$,
\begin{equation}
E=4U_{12}I_{3}\frac{1-2\frac{\pi\xi^2}{v_{c}}}
{1-\frac{3\pi\xi^2}{2v_{c}}}.
\end{equation}
Following the similar procedure performed in the previous section,
we find that when
\begin{equation}
\frac{\pi\xi^2}{v_{c}}>0.4 \label{trancondiTF}
\end{equation}
a transition of $r_{0}$ will occur as $t$ decreases. When the
condition (\ref{trancondiTF}) is satisfied, the critical value of
$t/U_{12}$ is
\begin{equation}
\frac{t}{U_{12}}=8I_{3}\sqrt{\frac{5\frac{\pi\xi^2}{v_{c}}-2}
{1-\frac{3\pi\xi^2}{2v_{c}}}}.
\end{equation}
When $t/U_{12}$ adiabatically decreases below this critical value,
$r_{0}$ will jump from zero to a non-zero value, and the
inter-layer quantum tunnelling will be suppressed immediately. The
phase diagram is shown in Fig(\ref{phasediagram3}). It is
physically similar to the right side of Fig.(\ref{phasediagram}),
but illustrated in an alternative way.

\section{Discussions and Conclusions}

So far we have studied the phase diagram of double-layer rotating
BEC in quantum Hall limit and Thomas-Fermi limit respectively. The
complexity as illustrated in above discussions once again
demonstrates the rich physics of rotating BEC and quantum
coherence effect. Our model contains rich physics due to the four
competing energies. The competition between intra-layer kinetic
energy, whose characteristic value is $\hbar\Omega$, and the
intra-layer interaction energy $U$ determines the properties of
vortex lattice. The competition between inter-layer kinetic
energy, whose characteristic value is $t$, and $U$ determines the
phase fluctuation between layers. The effect of the competition
between $t$ and $U_{12}$ is emphasized in this paper. However
these three effects are entangled together and interact strongly
with each other, that leads to the abundant physical phenomena.

What is most interesting is that under certain condition the
system will exhibit a new kind of quantum phase transition as
$t/U_{12}$ decreases. The transition is characterized by two sets
of vortex lattice being staggered, and consequently the loss of
phase coherence after the transition. Hence it presents a novel
mechanism for superfluid to Mott transition. Furthermore, the
condition that such a transition can happen depends on the
intrinsic properties of each condensate. In TFMF regime, the
condition is that the ratio of vortex core area to the area of
unit cell should be larger than a critical value. While in QHMF
regime the condition is that the lattice structure should be flat
enough. We remark that the different criteria reflect the
essential difference of these two physical regimes. Because how
much energy the density-density interaction term can gain in the
staggered lattices phase depends on the density undulation caused
by vortices, in the TFMF regime the density undulation is mostly
caused by vortex core structure. While in QHMF regime individual
vortices cores are smeared, the density and phase singularities
are strongly coupled, the density undulation therefore directly
relies on how vortices arrange themselves. This indicates that the
study of inter-layer coupling can be used as a powerful tool to
reveal the intra-layer physics.

There are still many opening interesting questions. So far we
still have no effective method to study the intermediate region
between TFMF regime and QHMF regime, which may be relevant to the
recent experiments of fast rotating BEC.\cite{CornellLLL} Besides,
if the rotating BEC is loaded into a one-dimensional optical trap,
it is possible to achieve the regime that the atom number in each
site is comparable to the vortex number, and in each layer the
Bose atoms are strongly correlated.\cite{Ho2} Such a system may
behave like a bosonic multi-layer fractional quantum Hall system.
The discussion made in this model can also be applied to a
rotating spin-$1/2$ condensate, with two hyperfine spin states
coupled by photons. However in that case $U_{12}$ may be
comparable with $U$, and the situation will be more complicated.
More fruitful physics is expected in further investigation, and we
hope that the theoretical results obtained in the present work
will stimulate more experiments.

\textit{Acknowledgement}\ \ HZ would like to thank Professor C. N.
Yang for his continuous encouragement and guidance. And the
authors would like to acknowledge Professor T. L. Ho and Z. Y.
Weng for helpful discussions. This work is supported by National
Natural Science Foundation of China ( Grant No. 10247002 ) and the
Ministry of Education of China.

\begin{appendix}
\section{Calculation of Coherent Hopping Terms in QHMF Regime\label{A1}}
In this appendix we will calculate the coherent hopping terms
between two layers with the help of the Jacobi theta function in
the QHMF regime. Using Eq.(\ref{theta function}) we have
\begin{eqnarray}
&&\theta^*(\zeta,\tau)\theta(\zeta+\bar{r}_{0},\tau)
\nonumber\\
&&=\sum\limits_{n_{1}=-\infty}^{+\infty}(-1)^{n_{1}}e^{-i\pi\tau^*(n_{1}+\frac{1}{2})^2+2\pi
i\zeta^* (n_{1}+\frac{1}{2})}\nonumber\\&&
\times\sum\limits_{n_{2}=-\infty}^{+\infty}(-1)^{n_{2}}e^{i\pi\tau(n_{2}+\frac{1}{2})^2+2\pi
i\zeta (n_{2}+\frac{1}{2})}\nonumber\\
&&=\sum\limits_{m_{1}}(-1)^m_{1}\exp[2\pi i
m_{1}\bar{x}-\frac{1}{2}v\pi
m_{1}^2+i\pi(\bar{x}_{0}+i\bar{y}_{0})m_{1}]L_{m_{1}},\nonumber\\
\end{eqnarray}
where $\bar{r}_{0}=r_{0}/b_{1}$, $m_{1}=n_{2}-n_{1}$ and
$m^\prime=n_{1}+n_{2}+1$.

Here $L_{m_{1}}$ is explicitly written as
\begin{eqnarray}
&&L_{m_{1}}=\sum\limits_{n_{1}=-\infty}^{+\infty}e^{i\pi
um_{1}m^{\prime}-2\pi\bar{y}m^\prime-\frac{v}{2}\pi m^{\prime
2}+\pi i
(\bar{x}_{0}+i\bar{y}_{0})m^{\prime}}\nonumber\\
&&=\sum\limits_{m_{2}}\int dn_{1}e^{i\pi
um_{1}m^{\prime}-2\pi\bar{y}m^\prime-\frac{v}{2}\pi m^{\prime
2}+\pi i (\bar{x}_{0}+i\bar{y}_{0})m^{\prime}}e^{2\pi i
m_{2}n_{1}}\nonumber\\
&&=\sum\limits_{m_{2}}(-1)^{(m_{1}+1)m_{2}}\frac{1}{2}\nonumber\\&&\times\int
dm^\prime e^{(i\pi um_{1}-2\pi\bar{y}+i\pi\bar{x}_{0}-\pi
\bar{y}_{0}+i\pi
m_{2})m^\prime-\frac{1}{2}v\pi m^{\prime 2}}\nonumber\\
&&=\sum\limits_{m_{2}}\frac{1}{\sqrt{2v}}(-1)^{(m_{1}+1)m_{2}}
e^{-\pi(m_{2}+\bar{x}_{0}+um_{1}+2i\bar{y}+i\bar{y}_{0})^2/2v}.\nonumber\\\label{Lm}
\end{eqnarray}
By defining
\begin{equation}
\vec{K}=\left[2\pi
m_{1}\hat{x}-2\pi\frac{m_{2}+um_{1}}{v}\hat{y}\right]\frac{1}{b_{1}}
\end{equation}
and
\begin{equation}
\vec{r}_{0}=\bar{x}_{0}\hat{x}+\bar{y}_{0}\hat{y},
\end{equation}
we can verify that
\begin{equation}
\pi\bar{x}_{0}m_{1}-\pi\bar{y}_{0}(m_{2}+um_{1})/v=\frac{1}{2}\vec{r}_{0}\cdot\vec{K}\label{rcdotk}
\end{equation}
and
\begin{equation}
-\pi\bar{y}_{0}m_{1}-\pi\bar{x}_{0}(m_{2}+um_{1})/v=\frac{1}{2}\vec{r}_{0}\times\vec{K}.\label{rcrossK}
\end{equation}
Using Eq.(\ref{Lm}),(\ref{rcdotk}) and (\ref{rcrossK}) we can
obtain that
\begin{eqnarray}
&&\theta^*(\zeta,\tau)\theta(\zeta+\bar{r}_{0},\tau)\nonumber\\
&&=\sum\limits_{K}f_{K}e^{i\vec{K}\cdot\vec{r}}e^{\frac{\pi}{v}
[2\bar{y}^2-2i\bar{y}(\bar{x}_{0}+i\bar{y}_{0})-(\bar{x}_{0}+i\bar{y}_{0})^2/2]},
\end{eqnarray}
where
\begin{equation}
f_{K}=\frac{(-1)^{m_{1}+m_{2}+m_{1}m_{2}}}{\sqrt{2v}}e^{\frac{1}{2}[(\vec{r}_{0}\times\vec{K})+i(\vec{r}_{0}\cdot\vec{K})]
-v_{c}|K|^2/(8\pi)}.
\end{equation}
Therefore
\begin{eqnarray}
&&\varphi_{1}^{*}\varphi_{2}\propto\sum\limits_{K}f_{K}e^{i\vec{K}\cdot\vec{r}}
\exp\left[-\frac{\pi}{2v}(2i\bar{y}+\bar{x}_{0}+i\bar{y}_{0})^2\right]\nonumber\\&&\times
\exp\left[\pi\frac{w^{*2}+(w+r_{0})^2}{2v_{c}}-\frac{r^2+(r+r_{0})^2}{2a_{\perp}^2}\right]\nonumber\\&&=
\sum\limits_{K}f_{K}e^{i\vec{K}\cdot\vec{r}}\nonumber\\&&\times\exp\left[-\frac{1}{\sigma^2}(r^2+\vec{r}\cdot
\vec{r}_{0})-\frac{r_{0}^2}{2a_\perp^2}+\frac{\pi}{v_{c}}i(\vec{r}\times\vec{r}_{0})\right]\label{coherenthopping}
\end{eqnarray}
and
\begin{eqnarray}
&&\varphi_{1}^{*
2}\varphi_{2}^2\propto\sum\limits_{KK^\prime}f_{K}f_{K^\prime}e^{i(\vec{K}+\vec{K}^\prime)\cdot\vec{r}}
\nonumber\\
&&\times \exp\left[-\frac{2}{\sigma^2}(r^2+r\cdot
r_{0})-\frac{r_{0}^2}{a_\perp^2}+\frac{2\pi}{v_{c}}i(\vec{r}\times\vec{r}_{0})\right].\nonumber
\end{eqnarray}

Then we can perform a integral over the two-dimensional space.
After considering the normalized conditions, it yields
%\begin{widetext}
\begin{eqnarray}
&&\int
dxdy\varphi^{*}_{1}\varphi_{2}=\nonumber\\
&&\frac{\sum\limits_{K}f_{K}
e^{(\frac{1}{4\sigma^2}-\frac{\pi\sigma^2}{v_{c}}\frac{\pi}{v_{c}}-\frac{1}{2a_{\perp}})r_{0}^2
+\frac{1}{2}\frac{\pi\sigma^2}{v_{c}}\vec{r_{o}}\times\vec{K}
-\frac{\sigma^2|K|^2}{4}+i\vec{K}\cdot\vec{r_{0}}}}{\sum\limits_{K}g_{K}
e^{-\frac{\sigma^2|K|^2}{4}}}.\nonumber\\\label{hopping term}
\end{eqnarray}
%\end{widetext}

\section{Calculation of Coherent Hopping Terms in the TFMF Regime
\label{A3}}

In this appendix we will calculate the energy of coherent hopping
term in the TFMF regime. Begin with Eq.(\ref{hopterm}) we have
\begin{eqnarray}
&& \Re \int d^2r
\varphi_{1}^{*}\varphi_{2}=\sum\limits_{r_{ij}}\int\limits_{\text{unit
cell}}d^2r
\sqrt{\rho_{1}\rho_{2}}\cos[\frac{\pi}{v_{c}}(\vec{r}\times\vec{r}_{0})]=\nonumber\\
&&\sum\limits_{r_{ij}}\int\limits_{\text{u.c}}d^2r\sqrt{\rho_{1}\rho_{2}}
\cos[\frac{\pi}{v_{c}}(\vec{r}-\vec{r_{ij}})\times\vec{r}_{0}]
\cos[\frac{\pi}{v_{c}}(\vec{r}_{ij}\times\vec{r}_{0})],\nonumber\\
\label{hoppingintegral1}
\end{eqnarray}
the $\sin$ term is cancelled out due to the spatial reflection
symmetry.

Approximately in each unit cell we take
\begin{equation}
\rho_{1}=\frac{\rho_{0}(r_{ij})}{1-\frac{\pi\xi^2}{v_{c}}}[1-\exp\left(-\frac{(r-r_{ij})^2}{\xi^2}\right)]
\end{equation}
and
\begin{equation}
\rho_{2}=\frac{\rho_{0}(r_{ij})}{1-\frac{\pi\xi^2}{v_{c}}}[1-\exp\left(-\frac{(r-r_{ij}-r_{0})^2}{\xi^2}\right)],
\end{equation}
hence
\begin{eqnarray}
&&\int\limits_{\text{u.c}}d^2r\sqrt{\rho_{1}\rho_{2}}
\cos[\frac{\pi}{v_{c}}(\vec{r}-\vec{r_{ij}})\times\vec{r}_{0}]\frac{1-\frac{\pi\xi^2}{v_{c}}}{\rho_{0}(r_{ij})}=\nonumber\\
&&\int\limits_{\text{u.c}}d^2r^\prime \sqrt{(1-e^{-\frac{r^{\prime
2}}{\xi^2}})(1-e^{-\frac{(r^\prime-r_{0})^2}{\xi^2}})}
\cos[\frac{\pi}{v_{c}}\vec{r}^\prime\times \vec{r}_{0}]=\nonumber\\
&&v_{c}\int\limits_{\text{u.c}}d^2r^\prime
\sqrt{(1-e^{-\frac{r^{\prime
2}v_{c}}{\xi^2}})(1-e^{-\frac{(r^\prime-r_{0})^2v_{c}}{\xi^2}})}
\cos[\pi\vec{r}^\prime\times \vec{r}_{0}].\nonumber\\
\label{hoppingintegral}
\end{eqnarray}
Here $r^\prime$ is defined as $r-r_{ij}$. The second equality
follows from a scaling that $r^\prime$ and $r_{0}$ are replaced by
$r^\prime/\sqrt{v_{c}}$ and $r_{0}/\sqrt{v_{c}}$. We notice that
for the determined lattice structure the integral in
Eq.(\ref{hoppingintegral}) only depends on $v_{c}/\xi^2$ and
$r_{0}$, which is denoted as $F$ and can be obtained numerically.

\begin{figure}[htbp]
\begin{center}
\includegraphics[width=2.3in]
{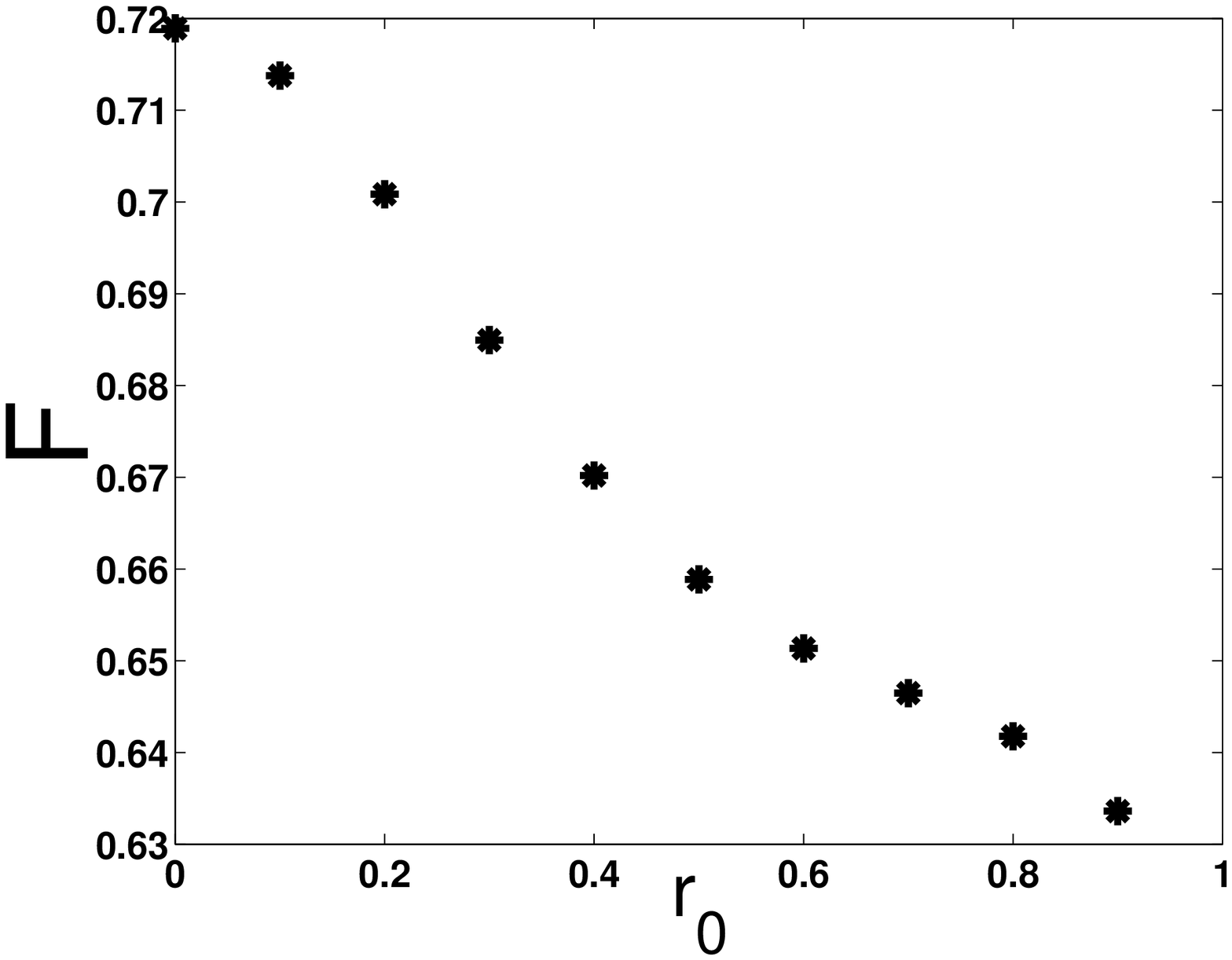}
\includegraphics[width=2.3in]
{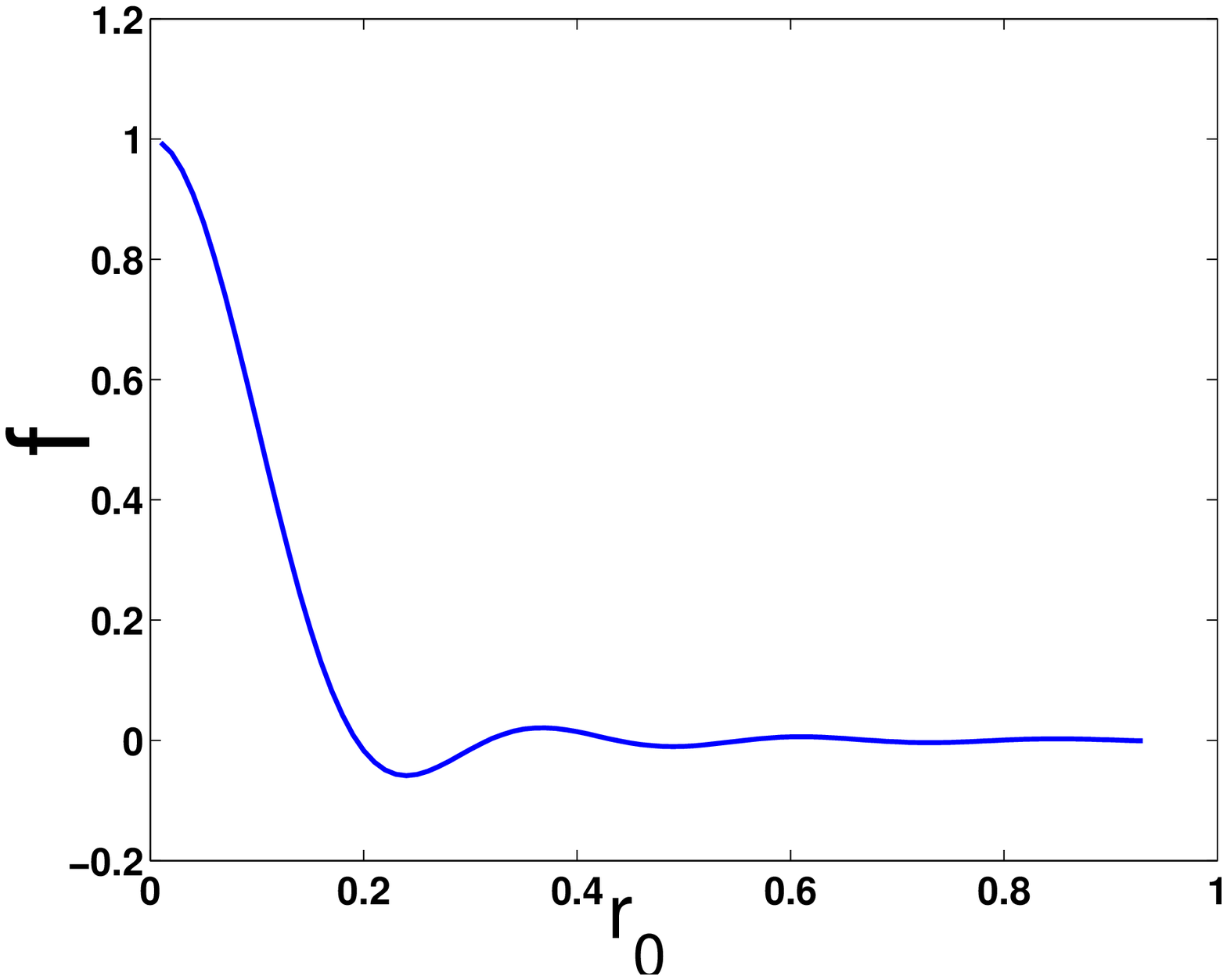}
 \caption{Upper: The integral $F$ as a function of $\frac{r_{0}}{\sqrt{v_{c}}}$.
 Lower: The integral $f$ as a function of
 $\frac{r_{0}}{\sqrt{v_{c}}}$. $v_{c}$
 is set to unity, $T=\pi\xi^2/v_{c}$ is 0.3 and $N_{v}$ is 225.    \label{Ffunction}}
\end{center}
\end{figure}

Therefore
\begin{eqnarray}
&&\text{Eq.(\ref{hoppingintegral1})}=
\frac{F(\frac{v_{c}}{\xi^2},\frac{r_{0}}{\sqrt{v_{c}}})}{1-\frac{\pi\xi^2}{v_{c}}}\sum\limits_{r_{ij}}\rho_{0}(r_{ij})v_{c}
\cos[\frac{\pi}{v_{c}}(\vec{r}_{ij}\times\vec{r}_{0})]\nonumber\\
&&\cong
\frac{F(\frac{v_{c}}{\xi^2},\frac{r_{0}}{\sqrt{v_{c}}})}{1-\frac{\pi\xi^2}{v_{c}}}\int
dr r d\theta \rho_{0}(r)
\cos[\frac{\pi}{v_{c}}r r_{0}\sin\theta]\nonumber\\
&&=2\pi \frac{
F(\frac{v_{c}}{\xi^2},\frac{r_{0}}{\sqrt{v_{c}}})}{1-\frac{\pi\xi^2}{v_{c}}}
\int dr r \rho_{0}(r)J_{0}(\frac{\pi}{v_{c}}rr_{0}),
\end{eqnarray}
where $J_{0}$ is Bessel function of zero order. At $r_{0}=0$, $F$
equals to $1-\pi\xi^2/v_{c}$. As $r_{0}$ increases $F$ will
decrease. The numerical result is shown in Fig.(\ref{Ffunction})

Besides, let $t=r/R$ with $R$ being the Thomas-Fermi radius we
have
\begin{eqnarray}
&&2\pi\int dr r
\rho_{0}(r)J_{0}(\frac{\pi}{v_{c}}rr_{0})\nonumber\\
&&=\frac{\pi}{2g}m(\omega^2-\Omega^2)R^4\int dt t
(1-t^2)J_{0}\left(\frac{\pi}{v_{c}}r_{0}Rt\right)\nonumber\\
&&=\frac{8}{(\pi r_{0} R)^2/v_{c}^2}J_{2}\left(\frac{\pi
r_{0}R}{v_{c}}\right).\label{BesselJ2}
\end{eqnarray}
The second equality follows from the Sonine integral identity and
the fact that $\rho_{0}$ is normalized. It yields unit at
$r_{0}=0$ and therefore Eq.(\ref{hoppingintegral1}) results in
unit as we expected. Notice that
\begin{equation}
\frac{\pi r_{0} R}{v_{c}}=\sqrt{\frac{\pi
R^2}{v_{c}}}\sqrt{\frac{\pi
r_{0}^2}{v_{c}}}=\sqrt{N_{v}}\sqrt{\frac{\pi r_{0}^2}{v_{c}}},
\end{equation}
where $N_{v}$ is the number of vortices contained in the cloud,
and its typical value is $100\sim 250$ in current experiments. We
choose $\sqrt{N_{v}}$ being $15$ in the following analysis, the
result of Eq.(\ref{BesselJ2}) is therefore
\begin{equation}
\text{Eq.(\ref{BesselJ2})}=\frac{8}{225\sqrt{\pi}}\frac{1}{r_{0}^2/v_{c}}J_{2}
\left(15\sqrt{\pi}\frac{r_{0}}{\sqrt{v_{c}}}\right).\label{f}
\end{equation}
Denoting the righthand side of Eq.(\ref{f}) by $f$, $f$ as a
function of $r_{0}/\sqrt{v_{c}}$ is plotted in the righthand side
of Fig.(\ref{Ffunction}). We find that $f$ is of the order
$10^{-4}$ and the hopping energy can be neglected when
$r_{0}=(b_{1}+b_{2})/2$.

In the same way we can calculate the two-particle hopping energy,
\begin{eqnarray}
&&\Re \int dxdy \varphi_{1}^{* 2}\varphi_{2}^2=\frac{2\pi
W(\frac{v_{c}}{\xi^2},\frac{r_{0}}{\sqrt{v_{c}}})}{\left(1-\frac{\pi\xi^2}{v_{c}}\right)^2}
\int dr r \rho_{0}^2
J_{0}\left(\frac{2\pi}{v_{c}}rr_{0}\right)\nonumber\\
&&=\frac{2\pi
W(\frac{v_{c}}{\xi^2},\frac{r_{0}}{\sqrt{v_{c}}})}{\left(1-\frac{\pi\xi^2}{v_{c}}\right)^2}
\frac{64}{\pi R^2(2\pi r_{0} R/v_{c})^3}J_{3}\left(\frac{2\pi
r_{0} R}{v_{c}}\right),\label{twohoppingTF}
\end{eqnarray}
where $W$ function is defined as the integral
\begin{equation}
\int\limits_{\text{u.c}} d^2r
(1-e^{-\frac{r^2\xi^2}{v_{c}}})(1-e^{-\frac{(r-r_{0})^2\xi^2}{v_{c}}})\cos\left(2\pi\vec{r}\times\vec{r}_{0}\right).
\end{equation}
Similarly when $r_{0}=(b_{1}+b_{2})/2$ Eq.(\ref{twohoppingTF}) is
about of the order $10^{-8}$ and need not be considered.

\end{appendix}


\begin{thebibliography}{99}
\bibitem{approximation}
The terms such as $\int u_{1}^3u_{2}$ are neglected in the energy
functional because they are smaller than the $U_{12}$ term and can
be effectively absorbed into the $t$ term.
\end{thebibliography}
\end{document}